\documentclass[aps,prl,twocolumn,showpacs,floatfix,superscriptaddress]{revtex4-1}
\usepackage{color}
\usepackage{amsfonts}
\usepackage[figuresright]{rotating}
\usepackage{amssymb}
\usepackage{amsmath}
\usepackage{psfrag}
\usepackage{subfigure}
\usepackage{multirow}
\usepackage{tabularx}
\usepackage{textcomp}
\usepackage{units}
\usepackage{hyperref}
\hypersetup{
 pdfnewwindow=true, colorlinks=true,
 linkcolor=blue, anchorcolor=blue,
 citecolor=blue, filecolor=blue,
 menucolor=blue, urlcolor=blue}

\usepackage{graphicx}
\usepackage{dcolumn}
\usepackage{bm}
\usepackage{color}

\makeatletter

\begin{document}
\title{Structural phase transition and material properties of few-layer monochalcogenides}

\author{Mehrshad\ \surname{Mehboudi}}
\affiliation{Department of Physics, University of Arkansas, Fayetteville, AR 72701, USA}

\author{Benjamin\ \surname{M. Fregoso}}
\affiliation{Department of Physics, University of California, Berkeley, CA, 94720, USA}

\author{Yurong\ \surname{Yang}}
\affiliation{Department of Physics, University of Arkansas, Fayetteville, AR 72701, USA}

\author{Wenjuan\ \surname{Zhu}}
\affiliation{Department of Electrical and Computer Engineering, University of Illinois at Urbana-Champaign, Urbana, IL 61801, USA}

\author{Arend\ \surname{van der Zande}}
\affiliation{Department of Mechanical Science and Engineering, University of Illinois at Urbana-Champaign, Urbana, IL 61801, USA}

\author{Jaime\ \surname{Ferrer}}
\affiliation{Departamento de F{\'\i}sica, Universidad de Oviedo, Asturias, Spain}

\author{L.\ \surname{Bellaiche}}
\affiliation{Department of Physics, University of Arkansas, Fayetteville, AR 72701, USA}

\author{Pradeep\ \surname{Kumar}}
\affiliation{Department of Physics, University of Arkansas, Fayetteville, AR 72701, USA}

\author{Salvador\ \surname{Barraza-Lopez}}
\email{sbarraza@uark.edu}
\affiliation{Department of Physics, University of Arkansas, Fayetteville, AR 72701, USA}

\begin{abstract}
GeSe and SnSe monochalcogenide monolayers and bilayers undergo a two-dimensional phase transition from a rectangular unit cell to a square unit cell at a temperature $T_c$ well below the melting point. Its consequences on material properties are studied within the framework of Car-Parrinello molecular dynamics and density-functional theory. No in-gap states develop as the structural transition takes place, so that these phase-change materials remain semiconducting below and above $T_c$. As the in-plane lattice transforms from a rectangle onto a square at $T_c$, the electronic, spin, optical, and piezo-electric properties dramatically depart from earlier predictions. Indeed, the $Y-$ and $X-$points in the Brillouin zone become effectively equivalent at $T_c$, leading to a symmetric electronic structure. The spin polarization at the conduction valley edge vanishes, and the hole conductivity must display an anomalous thermal increase at $T_c$. The linear optical absorption band edge must change its polarization as well, making this structural and electronic evolution verifiable by optical means.  Much excitement has been drawn by theoretical predictions of giant piezo-electricity and ferroelectricity in these materials, and we estimate a pyroelectric response of about $3\times 10^{-12}$ $C/K m$ here. These results uncover the fundamental role of temperature as a control knob for the physical properties of few-layer group-IV monochalcogenides.
\end{abstract}

\date{\today}

\maketitle

Few-layer group-IV monochalcogenides are semiconductors \cite{kaxiras,singh,zhu,gomes,li,benjamin,b2} with a structure similar to that of black phosphorus that exhibit a giant piezoelectric response in monolayer (ML) samples according to theory \cite{li,gomes3}. The four-fold degeneracy of their structural ground state first predicted by us in the past~\cite{usarXiv} leads to ferroelectricity \cite{usarXiv,Zeng,Qian}. These materials bring the concept of two-dimensional (2D) valleytronics on materials with reduced structural symmetries \cite{Pablo} closer to reality too \cite{gomes2}. Ferroelectrics must also exhibit a pyroelectric response, yet no theoretical description of this process has been provided for these 2D materials as of now.

\begin{figure*}[tb]
\begin{center}
\includegraphics[width=1.0\textwidth]{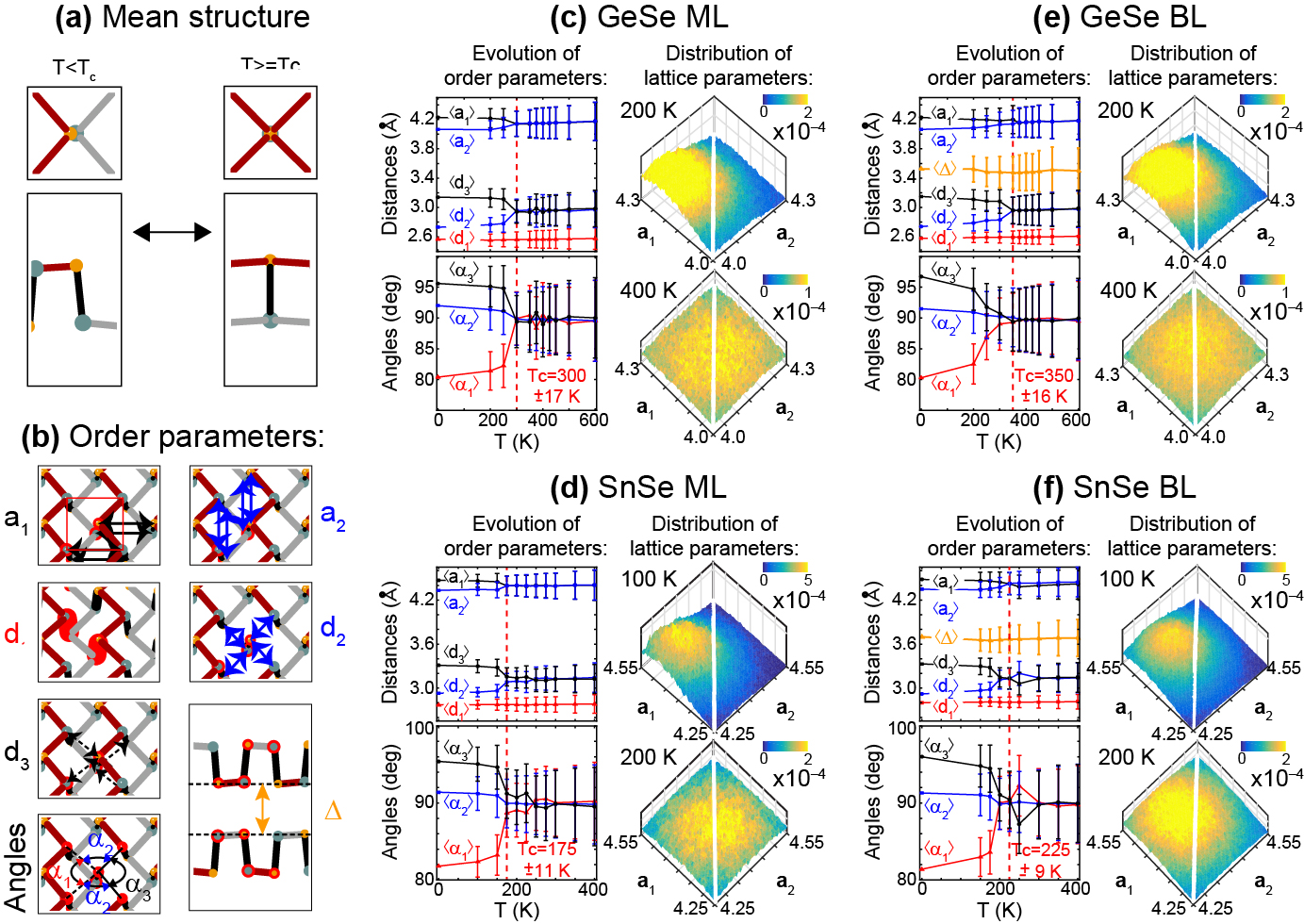}
\caption{(Color online.) (a) Schematic depiction of the structural transition. (b) Structural order parameters highlighting the transition.
(c to f) Left: thermal averages for the order parameters provided in (b) as a function of $T$ for GeSe and SnSe MLs and BLs. $T_c$ is reached when $\langle a_1\rangle= \langle a_2\rangle$, $\langle d_2\rangle =\langle d_3\rangle$, and $\langle\alpha_1\rangle=\langle\alpha_2\rangle=\langle\alpha_3\rangle$. The average distance among layers $\langle\Delta\rangle$ for BLs is shown too. Right: the distribution of lattice parameters $a_1$ and $a_2$ dramatically highlights the fluctuations leading to the error bars on the subplots on the left. The line $a_1=a_2$ is shown in white.}
\label{fig:fig1}
\end{center}
\end{figure*}

It remains unknown whether these materials undergo a complete degradation when exposed to air at the few-layer limit. Nevertheless, theory tells us that these monolayers host two-dimensional piezoelectricity, ferroelectricity, and a valley physics that is addressable with linearly-polarized light. Previous qualities do not exist simultaneously in any other known 2D atomic phase and justify additional theoretical and experimental studies. Adding to this list of properties, here we show that a structural transition taking place at finite temperature ($T$) modifies their band structure and hence their hole transport and optical properties, and induces a pyroelectric response. 
 Realizing these host of theoretical predictions requires thermally-controllable experimental studies of few-layer monochalcogenides in an inert atmosphere.

Theoretical results based on density-functional theory in Refs.~\cite{kaxiras,singh,zhu,gomes,gomes2,li,benjamin,b2,Zeng,Qian} correspond to structures at $T<T_c$ displayed in Fig.~\ref{fig:fig1}(a), and belong at the far left on the structure \emph{vs.} T plots in Figs.~\ref{fig:fig1}(c)-(f). We performed Car-Parrinello molecular dynamics (MD) calculations at finite $T$ \cite{siesta,CP,PR,Junquera,BH} on $8\times 8$ ML and AB-stacked bilayer (BL) supercells containing up to 512 atoms, with pseudopotentials and basis sets carefully validated \cite{PseudosPaper}, that led to the structural evolution at finite $T$ presented in Fig.~\ref{fig:fig1}(c-f). In order to focus on the results, thorough descriptions of methods, as well as the full time-evolution of the instantaneous $T$, total energy $E$, in-plane stress, and order parameters at selected target temperatures are provided as Supplemental Material (SM, Section I).

In Fig.~\ref{fig:fig1}(a) we illustrate a 2D structural transition whereby a rectangular unit cell with three-fold coordinated atoms at $T<T_c$ turns onto a square unit cell with five-fold coordinated atoms at $T\ge T_c$. The transition is captured in Figs.~\ref{fig:fig1}(c)-(f) by the thermal evolution of structural order parameters shown in Fig.~\ref{fig:fig1}(b) that include (i) lattice constants $a_1$ and $a_2$, obtained in four (eight) inequivalent ways in MLs (BLs) at any given unit cell, (ii) interatomic distances up to third nearest neighbors ($d_1$, $d_2$ and $d_3$), and (iii) angles subtended among a given atom and its second-nearest neighbors ($\alpha_3$), third-nearest neighbors ($\alpha_1$) and second- and third-nearest neighbor ($\alpha_2$).

The time auto-correlation of order parameters $a_1$ and $a_2$ --a measure of the time scale of structural fluctuations-- vanishes within  800 fs (Fig.~5, SM). Ensemble averages obtained from trajectories over 15,000 fs after thermal equilibration are reported in Figs.~\ref{fig:fig1}(c-f) for $\langle a_1\rangle$, $\langle a_2\rangle$, $\langle d_i\rangle$, and $\langle \alpha_i\rangle$ ($i=1,3$).

Sudden changes of structural order parameters signal the transition temperature $T_c$: $\langle a_1\rangle/\langle a_2\rangle>1$ at $T=0$ K, and  the transition is signaled by a rapid decay of $\langle a_1\rangle/\langle a_2\rangle$ onto unity. This ratio of lattice parameters decreases with increasing atomic number, so that SnSe MLs are expected to have a smaller $T_c$ than GeSe MLs \cite{usarXiv}. Additional signatures of the transition are the coalescence of $d_2$ and $d_3$ onto a single value, and the coalescence of in-plane angles defined in Fig.~\ref{fig:fig1}(b) toward 90$^o$. As seen in Fig.~\ref{fig:fig1}, the transition occurs at $T_c= 175\pm 11$ K for SnSe MLs and at a higher temperature of $350\pm 16$ K for GeSe MLs. It is interesting to note that the square unit cell --corresponding to a point of unstable equilibrium at $T=$0 K \cite{delaire,usarXiv}-- becomes, on average, the preferred structure at $T_c$.

Now, $\langle a_1\rangle/\langle a_2\rangle$ is known to increase with the number of layers for a given layered monochalcogenide as well \cite{gomes} and, accordingly, one should expect an increase of $T_c$ for a given material in going from MLs to BLs. Within the temperature resolution of 25 K employed in our calculations, we see a 50 K increase of $T_c$ in going from MLs to BLs. Such increase makes our results consistent with experiments on bulk SnSe, where $T_c$ is of the order of 800 K \cite{r1,dravid,delaire,Heremans1} (MD simulations of bulk samples require inclusion of four monolayers and are out of our reach).  The structural change discussed on this and previous paragraph should be experimentally observable with XRD techniques.

\begin{figure*}[tb]
\begin{center}
\includegraphics[width=1.0\textwidth]{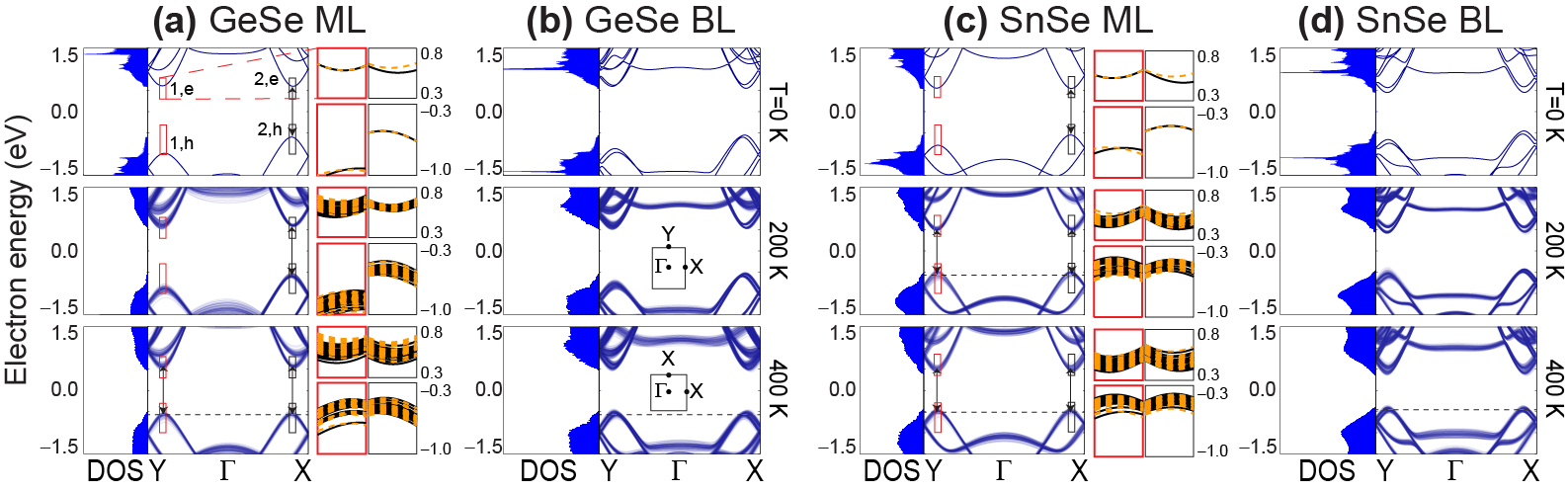}
\caption{(Color online.) Electronic DOS and band structures for $t_i=5000+100i$ fs ($i=1,2,3,...,150$) for (a) GeSe ML, (b) GeSe BL, (c) SnSe ML, and (d) SnSe BL at 0, 200, and 400 K. The DOS and band structures become broader with increasing $T$, but no in-gap states are seen in the DOS nor the band structures despite of fluctuations. The $X-$ and $Y-$points are inequivalent for $T<T_c$ and become equivalent for $T\ge T_c$ as $a_1=a_2$. The zero energy was set at the mid-gap. The thermal dependence of the hole conductivity should display an anomalous behavior at $T_c$ when valleys $1,h$ and $2,h$ align.}
\label{fig:fig2}
\end{center}
\end{figure*}

We note that a melting transition would be signaled by an isotropic increase of interatomic distances $\langle d_1\rangle$, $\langle d_2\rangle$ and $\langle d_3\rangle$. But the mean (inter-sublayer) distance $\langle d_1 \rangle$ in Figs.~\ref{fig:fig1}(c)-(f) remains constant through the transition, displaying smaller fluctuations than (intra-sublayer) distances $\langle d_2 \rangle$ and $\langle d_3\rangle$, so that individual MLs retain their 2D character through the transition. An additional (geometrical) argument for the 2D character of the transition can be made from $\langle \alpha_i\rangle$ ($i=1,2,3$) too: $\langle\alpha_1\rangle+2\langle\alpha_2\rangle+\langle\alpha_3\rangle$ add up to $2\pi$. The angle defect, defined as $2\pi-(\langle\alpha_1\rangle+2\langle\alpha_2\rangle+\langle\alpha_3\rangle)$, is equal to zero only on a planar structure \cite{ACSNano}.

Structural degeneracies lead to an \emph{anharmonic elastic energy profile} \cite{usarXiv} and hence to soft (``floppy'') phonon modes on monochalcogenide layered materials \cite{dravid,delaire,Heremans1}; anharmonicity makes it relevant to discuss fluctuations. The distribution of lattice parameters shown at the right subplots in Figs.~\ref{fig:fig1}(c)-(f) for increasing $T$ has a mean value converging towards the (white) diagonal line $a_1=a_2$ at $T_c$, which is consistent with a displacive transition \cite{delaire}. The maximum height of the distribution decays sharply nevertheless, making the distribution extremely broad as temperature raises. This broad distribution sets the error bar on $\langle a_1\rangle$ and $\langle a_2\rangle$ and is a signature of atomistic fluctuations (disorder). Excursions of $a_1$ and $a_2$ towards the right of the white $a_1=a_2$ line gain a finite probability with increasing temperature, and $a_1$ and $a_2$ have a rather homogeneous distribution at $T_c$: this distribution highlights the fluctuations of the order parameter. Considering these fluctuations, material properties to be discussed next were evenly sampled out of one hundred and fifty individual frames at times $t_i=5000+100i$ fs ($i=1,2,3,...,150$).

These materials remain semiconducting through the transition: the electronic density of states (DOS) obtained on instantaneous supercells at times $t_i$ in Fig.~\ref{fig:fig2} shows a well-defined bandgap for $T$ below and above $T_c$ (details of DOS calculations are disclosed in SM). The bandgaps in Fig.~\ref{fig:fig2} --whose magnitudes are explicitly reported in Table~I (SM)-- change by about 200 meV at 400 K with respect to their values at 0 K. The DOS has two additional features: (i) the sharpest peaks at 0 K that become blurred at increasing $T$ and (ii) shallow DOS pockets around the valence-band maximum for $T<T_c$.

\begin{figure*}[tb]
\begin{center}
\includegraphics[width=1.0\textwidth]{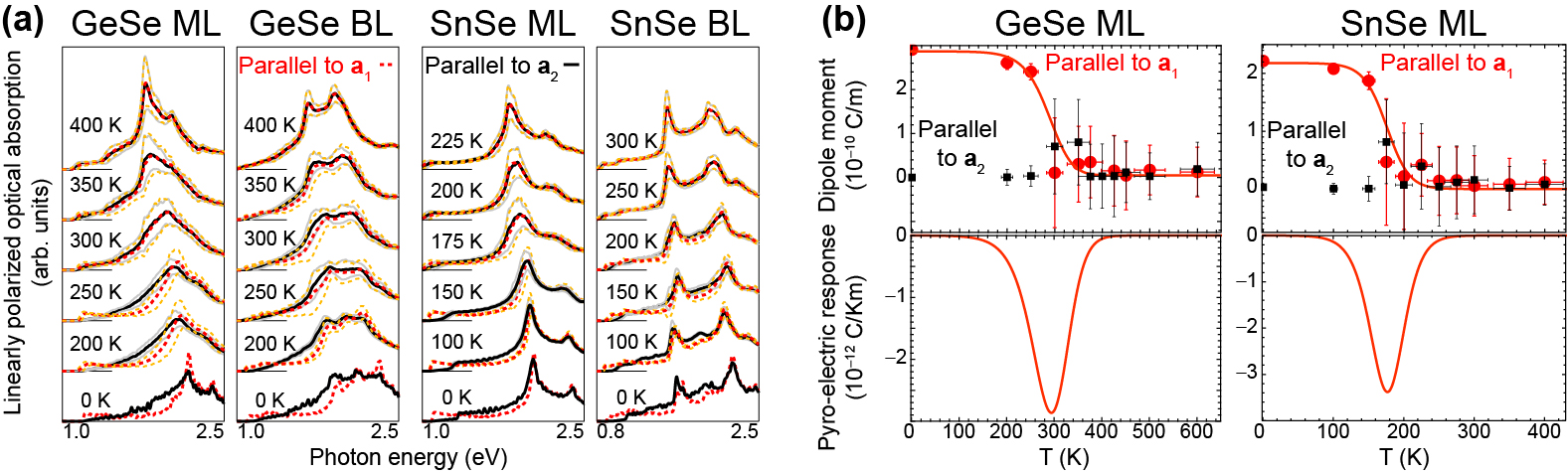}
\caption{(Color online.) (a) Linearly-polarized optical absorption spectra. The gray and dashed orange lines are error bars. (b) Upper subplots: thermal evolution of the electric dipole moment per unit cell $p$. Lower subplot: pyroelectric response $dp/dT$.}
\label{fig:fig3}
\end{center}
\end{figure*}

The band structures in Fig.~\ref{fig:fig2} were obtained from instantaneous unit cells built from average lattice and basis vectors at times $t_i$ defined above. The width of these bands is another experimentally-observable indicator of structural fluctuations that must be visible on ARPES data. The sharp peaks in the DOS at 0 K correspond to relatively flat bands located around the $\Gamma-$point whose energy location fluctuates with increasing $T$, thus making these peaks shallower. A band unfolding scheme \cite{purdue,laurent2,Ozaki,gollum} confirms these findings.

2D materials with reduced structural symmetries originate a novel paradigm in valleytronics in which crystal momentum labels individual valleys one-to-one \cite{Pablo}. In SnSe and GeSe MLs and BLs, the shallow DOS pocket at 0 K ($\langle a_1\rangle > \langle a_2\rangle$) corresponds with a hole-valley ($2,h$) located along the $\Gamma-X$ line in Fig.~\ref{fig:fig2} \cite{kaxiras,singh,gomes} that lies at a higher energy when contrasted to the hole-valley at the $\Gamma-Y$ line ($1,h$).

Band structure insets in Fig.~\ref{fig:fig2} show the effect of $T$ on valley spin polarization that arises due to spin-orbit coupling (SOC) \cite{soc}. The spin-polarization at these insets becomes drastically degraded at $T_c$ because spin-up (solid black) and spin-down (dashed yellow) bands become broader and closer together. For this reason, the remaining results on this Letter will neglect the effect of SOC. (AB BLs have zero spin polarizations at individual valleys due to inversion symmetry.)

As shown thus far, MLs and BLs increase their structural symmetry as $T_c$ is approached from below (Fig.~\ref{fig:fig1}). This means that  the $X-$ and $Y-$points in reciprocal space --which were inequivalent for $T<T_c$-- become equivalent for $T\ge T_c$ as $\langle a_1\rangle=\langle a_2\rangle$. As $T_c$ is reached, the hole valley along the $\Gamma-Y$ direction raises up to align with the valley located along the $\Gamma-X$ line (Fig.~2). One valley contributes to the hole conductivity at the band edge for $T<T_c$, while two valleys contribute at $T\ge T_c$, giving rise to an anomalous thermal dependence of the hole conductivity at $T=T_c$ that should be visible in standard transport measurements of gated or hole-doped samples.

As seen in Fig.~\ref{fig:fig3}(a), crystal momentum couples to the orientation of adsorbed linearly polarized light \cite{gomes2}. But the induced equivalence among the $X-$ and $Y-$points for $T\ge T_c$ makes the optical adsorption band edges for horizontally- and vertically-polarized light identical, making the band edge polarized at 45$^{o}$, which then represents an additional optical signature of the structural transition.

The binary composition of MLs and the asymmetry upon inversion about an axis parallel to $\mathbf{a}_2$ originates a net electric dipole $p$ along the longest lattice vector ($\mathbf{a}_1$) \cite{zhu}, resulting in a piezoelectric response at 0 K \cite{li,gomes3}. But as $\alpha_1$ $\alpha_2$, and $\alpha_3$ fluctuate (Fig.~\ref{fig:fig1}(c-f)), the orientation of these dipoles randomizes at $T_c$, turning the net electric dipole moment to zero. This hypothesis is demonstrated in Fig.~\ref{fig:fig3}(b) by averaging the mean electric dipole moment \cite{vanderbilt,vasp,PBE,paw1,paw2} over times $t_i$ at a given $T$ on instantaneous average unit cells (section VI, SM).

The three salient features of ferroelectrics are: (i) piezoelectricity, (ii) ferroelectricity, and (iii) pyroelectricity. The abrupt decay of $p$ around $T_c$ was fitted to sigmoidal functions, whose temperature derivative $dp/dT$ is the pyroelectric response given at lower subplots in Fig.~\ref{fig:fig3}(b). The pyroelectricity hereby predicted may very well be a first within the field of 2D atomic materials.

To conclude, we predict a structural transition in MLs and AB BLs of GeSe and SnSe. The transitions should be observable on mean values of lattice parameters and (in-plane) distances and angles among second and third nearest neighbors (XRD). These materials remain semiconductors through the transition, which should also be observable through ARPES, hole conductivity, and optical absorption measurements. We contributed the pyroelectric response of GeSe and SnSe MLs as well. These theoretical results may motivate and guide future experimental work in these few-layer materials with detailed thermal control and performed on an inert atmosphere.

M.M. and S.B.-L. are funded by an Early Career Grant from the DOE (Grant DE-SC0016139). Y. Y. and L. B. were funded by ONR Grant N00014-12-1-1034, and BMF by NSF DMR-1206515 and CONACyT (Mexico). J.F. acknowledges funding from the Spanish MICINN, Grant FIS2012-34858, and European Commission FP7 ITN MOLESCO (Grant 606728). Calculations were performed at Arkansas' {\em Trestles}.

\section{Supplementary material}

\subsection{Details of molecular dynamics calculations}

GeSe and SnSe monolayers (MLs) and bilayers (BLs) were studied with NPT Carr-Parrinello molecular dynamics (MD) runs on a fixed number of particles at finite temperature with the {\em SIESTA} DFT code. Calculations were performed on 8$\times$8 supercells, for up to 20,000 fs, with a 1.5 fs time resolution, and employed standard basis sets, pseudopotentials with van der Waals corrections of the Berland-Per Hyldgaard type.

\begin{figure}[tb]
\begin{center}
\includegraphics[width=0.48\textwidth]{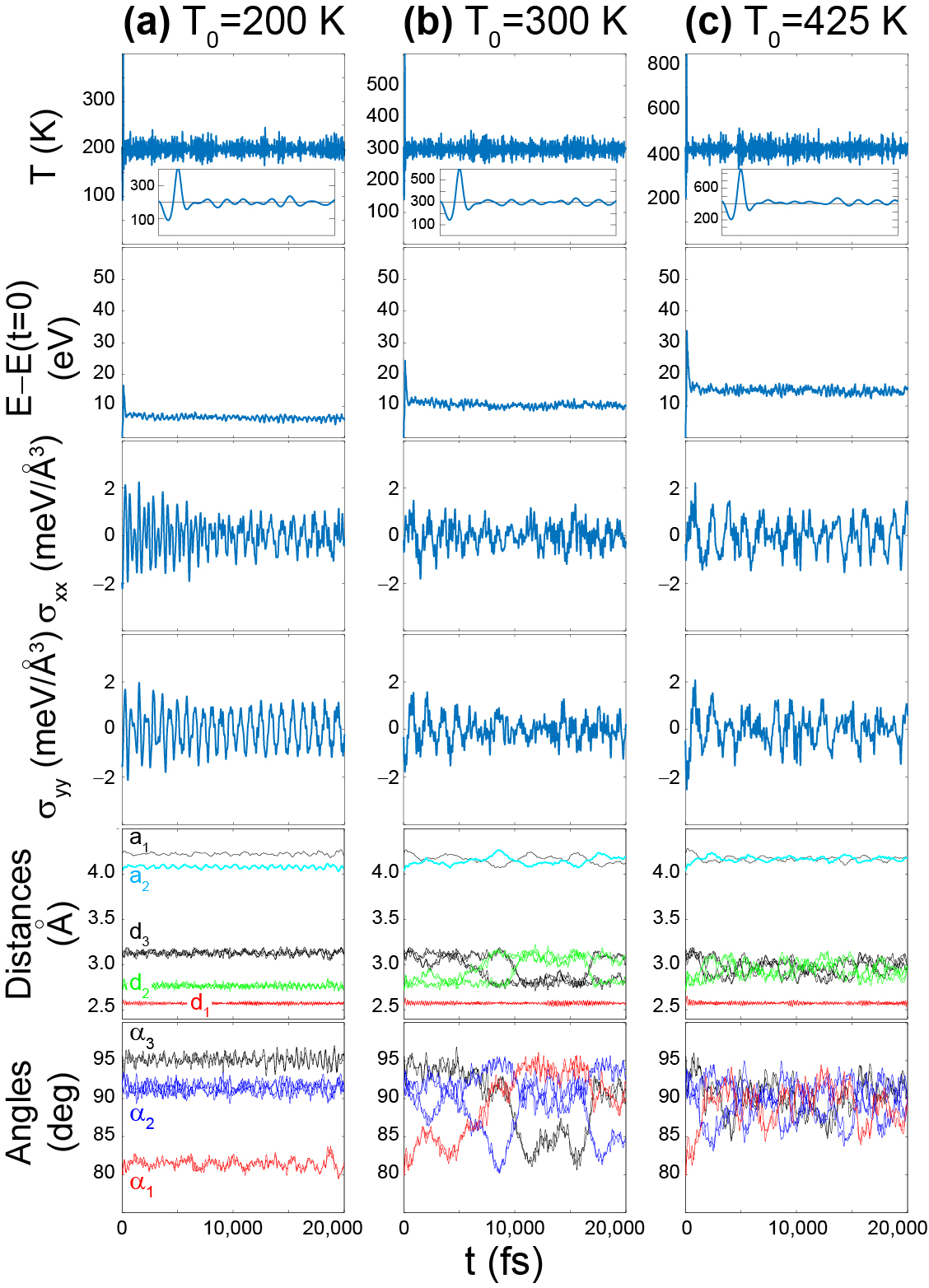}
\caption{(i) Instantaneous temperature $T$, (ii) Kohn-Sham supercell energy $E$, (iii) $\sigma_{xx}$, (iv) $\sigma_{yy}$, (v) $a_1$ (black), $a_2$ (blue), $d_1$ (red), $d_2$ (green), and $d_3$ (black), and (vi) angles $\alpha_1$ (red), $\alpha_2$ (blue), and $\alpha_3$ (black) as a function of time for a GeSe ML. Diagrams indicating atoms involved in estimating $a_1$, $a_2$, $d_1$, $d_2$, $d_3$, $\alpha_1$, $\alpha_2$ and $\alpha_3$ can be found in Figure 1(a) of the main text. Insets on instantaneous temperature $T$ subplots show thermal evolution over the first 1,500 fs.}
\label{fig:fig1si}
\end{center}
\end{figure}

Timely execution of MD calculations requires an appropriate choice of running parameters. For a fixed number of atoms, basis set, and time resolution for the MD calculation, the crucial parameters determining speed are: (a) the $k-$point sampling, (b) the target precision of the self-consistent electronic density cycle, and (c) the fineness of the real-space grid on which the Poisson equation is solved.

\begin{figure}[h]
\begin{center}
\includegraphics[width=0.48\textwidth]{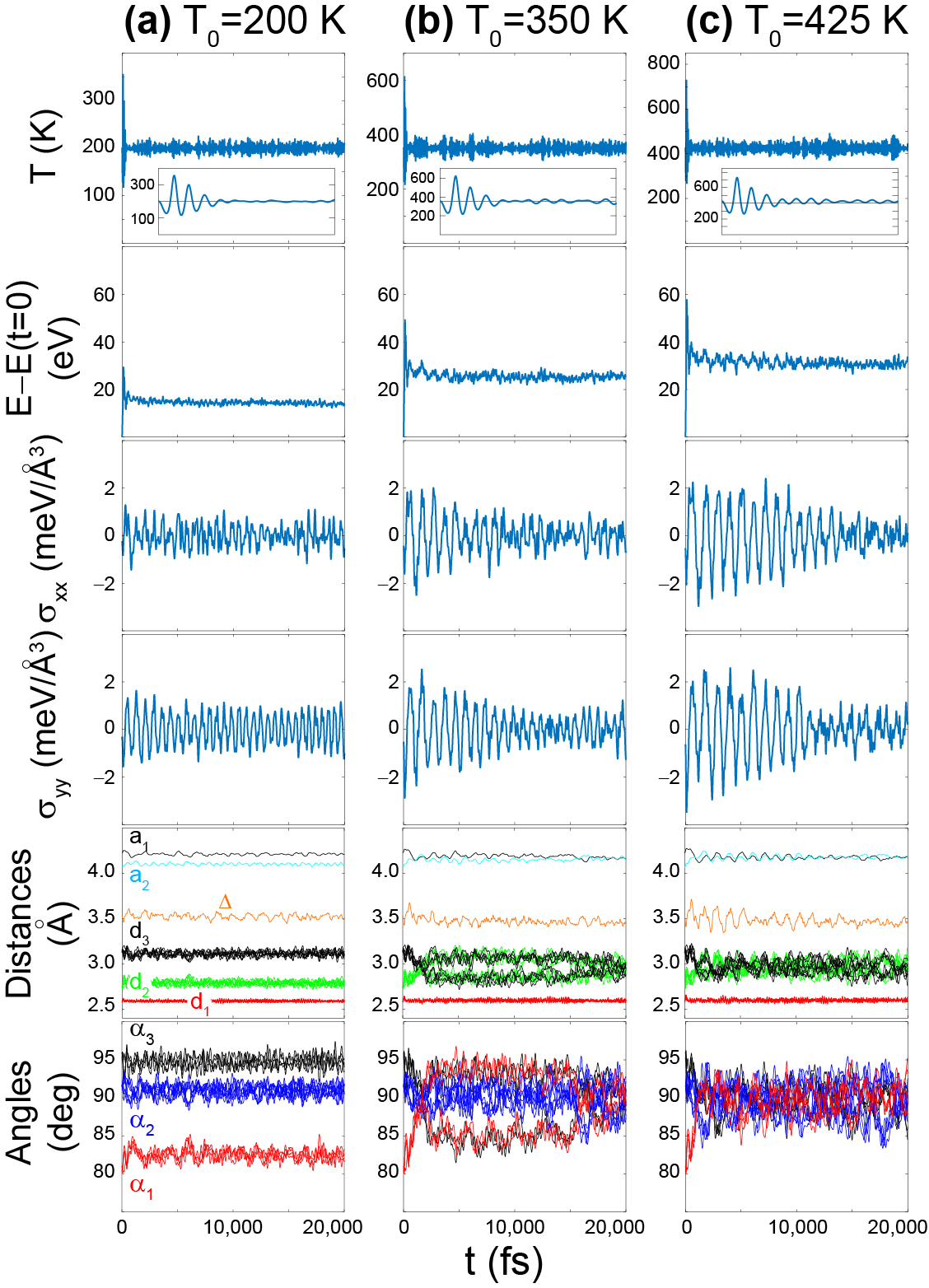}
\caption{(i) $T$, (ii) $E$, $\sigma_{xx}$, (iv) $\sigma_{yy}$, (v) $a_1$ (black), $a_2$ (blue), $\Delta$ (orange), $d_1$ (red), $d_2$ (green), and $d_3$ (black), and (vi) angles $\alpha_1$ (red), $\alpha_2$ (blue), and $\alpha_3$ (black) as a function of time for a GeSe BL.  Insets on instantaneous temperature $T$ subplots show thermal evolution over the first 1,500 fs.}
\label{fig:fig2si}
\end{center}
\end{figure}

We track the temperature $T(t)$, the total Kohn Sham energy $E(t)$, the in-plane components of the stress tensor $\sigma_{xx}$ and $\sigma_{yy}$, and structural parameters that include lattice constants, interatomic distances, and angles as a function of time $t$ (see Figs.~4 through 7, and Fig.~1(a) for graphical definitions).

Our initial choice of relevant input parameters included: (a) a $2\times 2\times 1$ $k-$point sampling of these $8\times 8$ supercells, (b) a standard precision of the electronic density equal to 10$^{-4}$, and (c) a real space grid with a cutoff of 300 Ry. We call these ``high-precision calculations,'' which were ran at $T=300$ K for GeSe and SnSe MLs and BLs.

We also ran calculations at $T=300$ K for up to 3000 fs using the following parameters: (a) a $1\times 1\times 1$ $k-$point sampling of the $8\times 8$ supercells (i.e., a sampling that only includes the $\Gamma-$point), (b) a precision of the electronic density equal to $5\times 10^{-4}$, and (c) a real space grid with a cutoff of 150 Ry. We call these ``normal precision calculations.''

We determined the sensitivity of the MD algorithm to minute changes on interatomic (Hellmann-Feynman) forces arising from these two choices of input parameters. In test runs for up to 1,000 fs and $T=300$ K, we find that $T(t)$, $E(t)$, $\sigma_{xx}(t)$ and $\sigma_{yy}(t)$, lattice constants, interatomic distances, and angles overlap when using both sets of input parameters: runs employing ``normal'' parameters display an evolution of all relevant physical parameters that are essentially identical to those from runs with ``high precision'' parameters. Nevertheless, they accrue significantly smaller walltimes that at the end make a detailed temperature sampling possible within our computing constraints. The main results reported in this work were obtained with MD runs that employed ``normal precision'' parameters.

\begin{figure}[tb]
\begin{center}
\includegraphics[width=0.48\textwidth]{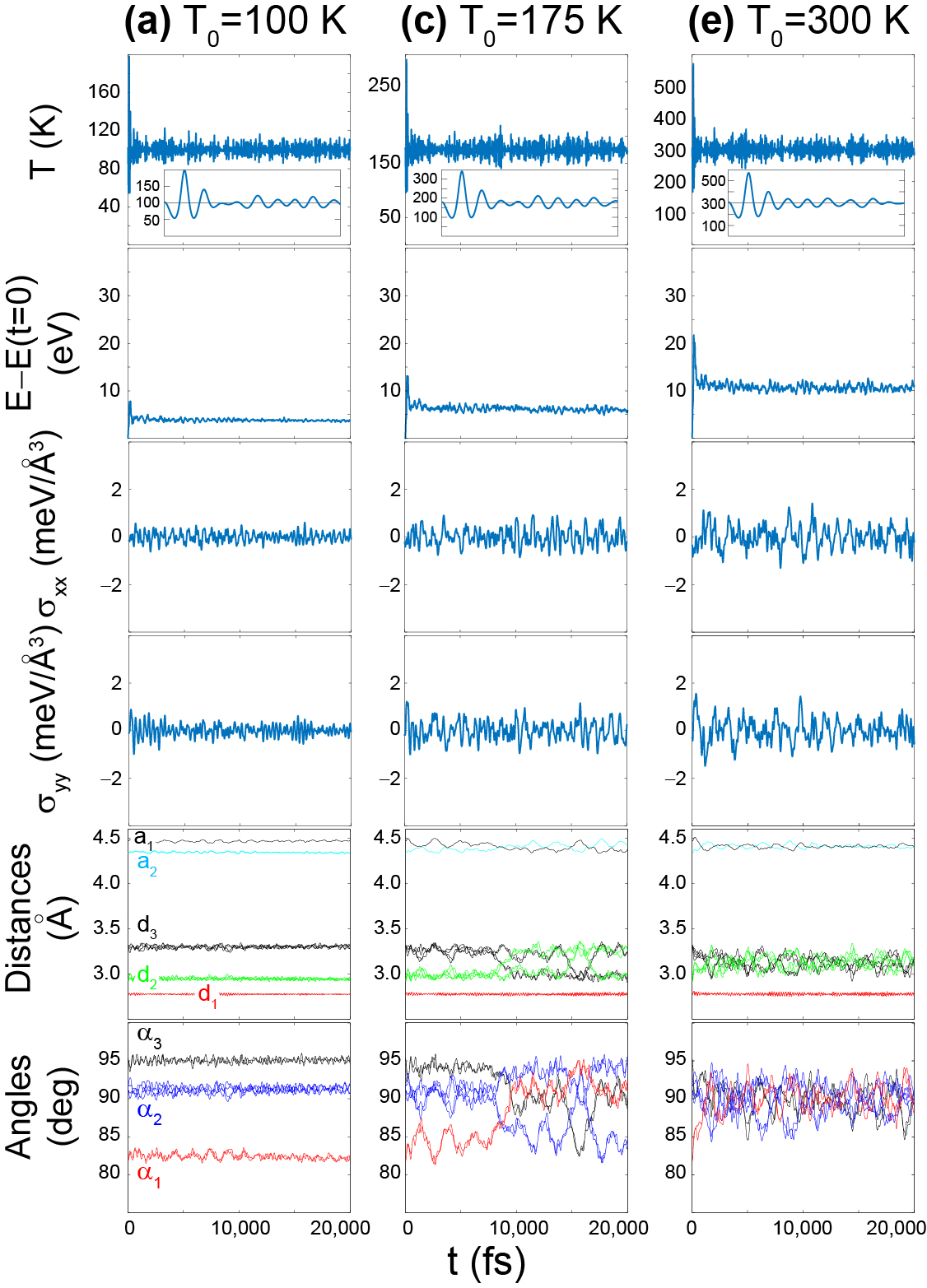}
\caption{(i) $T$, (ii) $E$, (iii) $\sigma_{xx}$, (iv) $\sigma_{yy}$, (v) $a_1$ (black), $a_2$ (blue), $d_1$ (red), $d_2$ (green), and $d_3$ (black), and (vi) $\alpha_1$ (red), $\alpha_2$ (blue), and $\alpha_3$ (black) as a function of time for a SnSe ML.  Insets on instantaneous temperature $T$ subplots show thermal evolution over the first 1,500 fs.}
\label{fig:fig3si}
\end{center}
\end{figure}

\subsection{Time auto-correlation of lattice parameters}
We next determine the time auto-correlation function, which applied to the lattice constants $a_i$ ($i=1,2$) has the following form:
\begin{equation}
C(t)\equiv \frac{\langle a_i^j(t)a_i^j(t=0)\rangle-\langle a_i\rangle^2}{\langle a_i^2\rangle-\langle a_i\rangle^2}.
\end{equation}
As shown in Fig.~8, the 2D materials studied here display a vanishing time auto-correlation of about 800 fs.

\begin{figure}[tb]
\begin{center}
\includegraphics[width=0.49\textwidth]{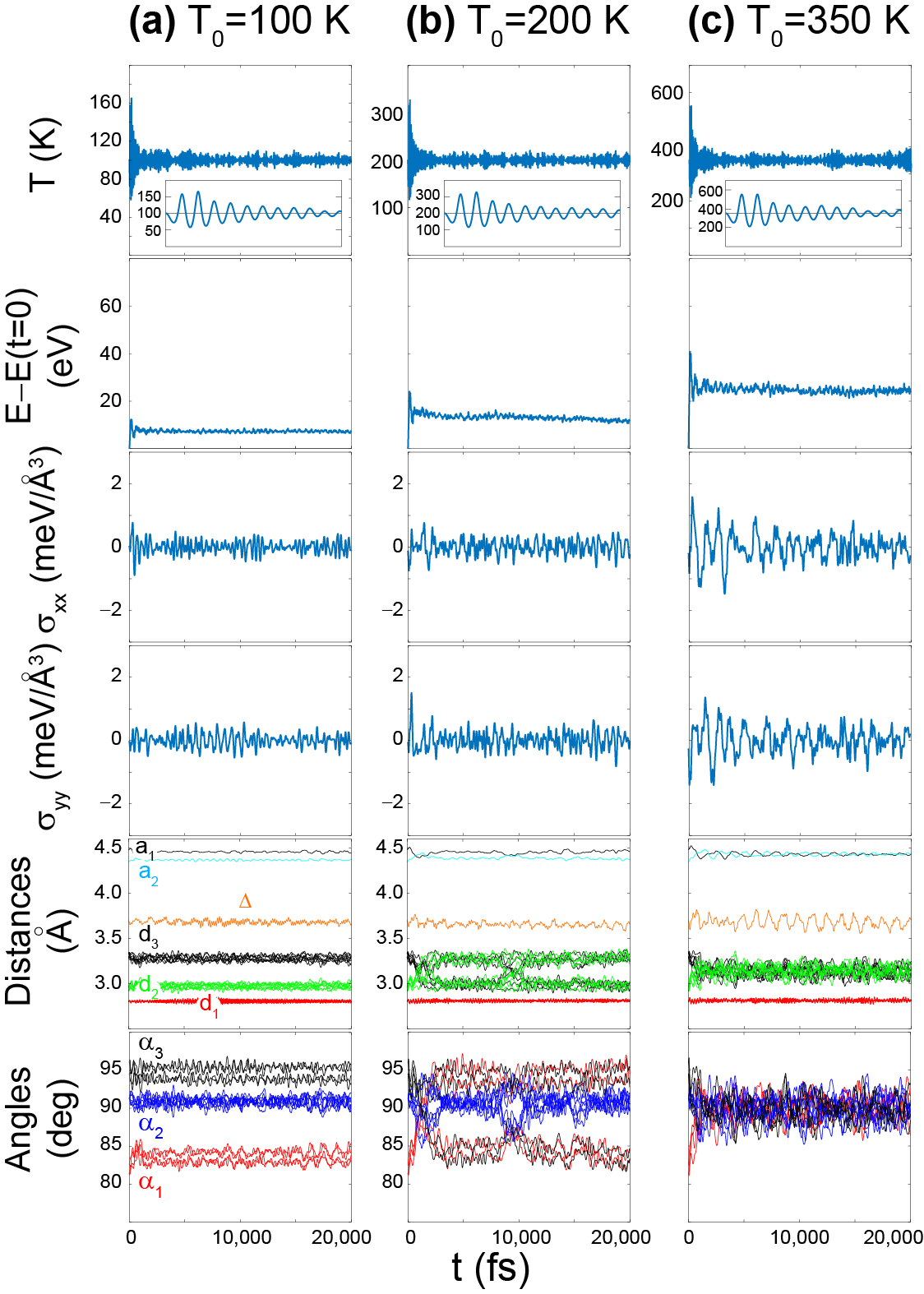}
\caption{(i) $T$, (ii) $E$, (iii) $\sigma_{xx}$, (iv) $\sigma_{yy}$, (v) $a_1$ (black), $a_2$ (blue), $\Delta$ (orange), $d_1$ (red), $d_2$ (green), and $d_3$ (black), and (vi) $\alpha_1$ (red), $\alpha_2$ (blue), and $\alpha_3$ (black) as a function of time for a SnSe BL.  Insets on instantaneous temperature $T$ subplots show thermal evolution over the first 1,500 fs.}
\label{fig:fig4si}
\end{center}
\end{figure}

\begin{figure}[tb]
\begin{center}
\includegraphics[width=0.49\textwidth]{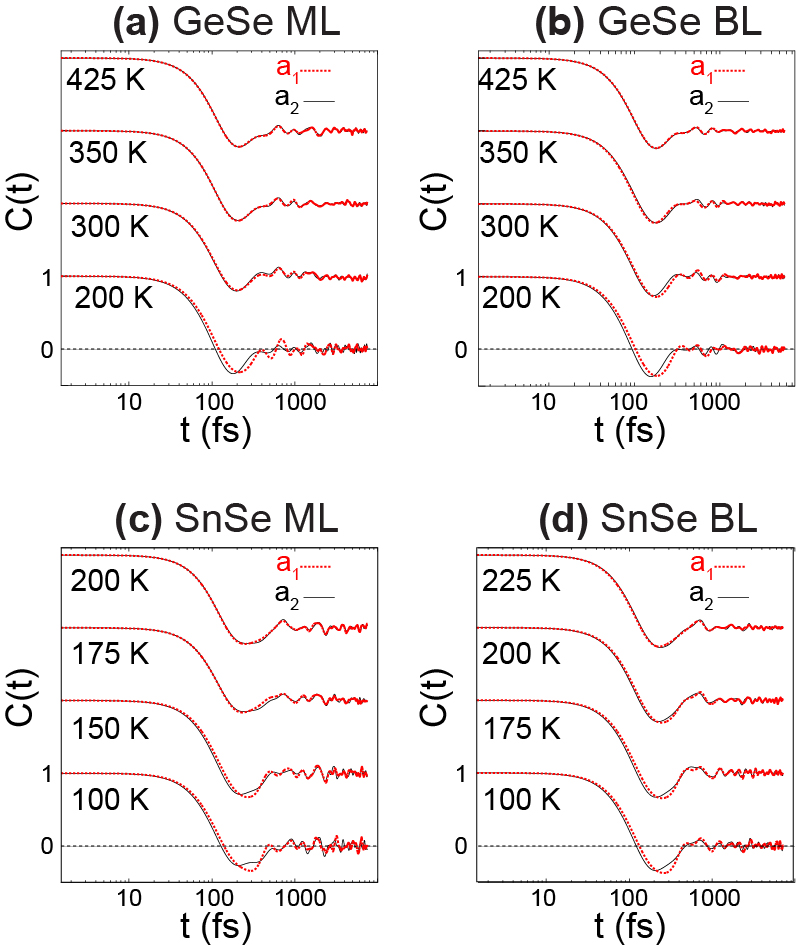}
\caption{The time auto-correlation of lattice parameters $a_1$ (dashed red) and $a_2$ (solid black) at varying temperatures vanishes within 800 fs.}
\label{fig:fig4si}
\end{center}
\end{figure}

\begin{table}[tb]
\caption{DFT electronic bandgap (in meV).}
\begin{tabular}{|c|ccc|}
\hline
\hline
T (K) &0  &200  &400 \\
\hline
GeSe ML	&	1195$\pm$0    &1020$\pm$53	&949$\pm$43	\\
GeSe BL	&	1104$\pm$0    &1076$\pm$22	&1005$\pm$40	\\
\hline
SnSe ML &   1096$\pm$0   &915$\pm$30	&854$\pm$38	\\
SnSe BL &   909$\pm$0    &874$\pm$25 &797$\pm$39	   \\
\hline
\hline
\end{tabular}
\label{table:table2}
\end{table}

\begin{table}[tb]
\caption{DFT energies (in meV) for direct optical transitions for the valley located along the $\Gamma-X$ high-symmetry line. These energies are consistent with the band edges reported in Figure 3 (see energy of adsorption edges parallel to $\mathbf{a}_1$), and are slightly larger than the indirect band-gaps reported in Table I.}
\begin{tabular}{|c|ccc|}
\hline
\hline
T (K) &0 &200  &400 \\
\hline
GeSe ML	&	1195$\pm$0   	&1115$\pm$33	&1066$\pm$32	\\
GeSe BL	&	1104$\pm$0   	&1083$\pm$15	&1049$\pm$41\\
\hline
SnSe ML &   1329$\pm$0  	&957$\pm$37	&931$\pm$35	\\
SnSe BL &   910$\pm$0   	&893$\pm$19	    &835$\pm$28	    \\
\hline
\hline
\end{tabular}
\label{table:table2}
\end{table}

\subsection{Details of DOS calculations and Evolution of the electronic bandgaps with temperature}

DOS calculations were performed on supercell structures at times $t_i$ defined in the main text; we employed a $k-point$ sampling of 30$\times$30$\times$1, a tolerance of the electronic density of $10^{-4}$, a Mesh cutoff of 300 Ry, and an energy resolution of 0.01 eV. The overall sum of individual DOS traces is shown as area DOS plots in Figure 2. Band structure plots were created from the instantaneous supercell averages of lattice parameters and basis atoms at times $t_i$ as well. There is a small fluctuation of the horizontal axis on bands plots that is a consequence of fluctuations of $a_1$ and $a_2$ that is neglected in Fig.~2 for simplicity.

An unfolding scheme developed for DFT calculations based on numerical atomic orbitals confirmed the band structures displayed on Fig.~2.

The numerical magnitude of the DFT bandgaps and energies for direct optical transitions at the valley band edges are reported in Tables I to III.
The valence band of the valley located along the $\Gamma-Y$ symmetry line moves upward in energy as $a_1$ and $a_2$ become equal in magnitude, making that transition energy decrease in a sudden manner for $T\ge T_c$.

The bands that include the effect of spin-orbit coupling, shown as insets to Fig.~2 were obtained with the Oviedo version of the SIESTA code.

\begin{table}[tb]
\caption{DFT energies (in meV) for direct optical transitions for the valley located along the $\Gamma-Y$ high-symmetry line. These energies are consistent with the band edges reported in Figure 3 (see energy of adsorption edges parallel to $\mathbf{a}_2$ and dashed vertical lines that serve as guides to the eye), and are slightly larger than the indirect band-gaps reported in Table I.}
\begin{tabular}{|c|ccc|}
\hline
\hline
T (K) &0  &200  &400 \\
\hline
GeSe ML	&	1597$\pm$0    &1318$\pm$89	&1036$\pm$61	\\
GeSe BL	&	1469$\pm$0    &1325$\pm$45	&1044$\pm$40 \\
\hline
SnSe ML &   1096$\pm$0    &947$\pm$64	&905$\pm$36	\\
SnSe BL &   1112$\pm$0    &958$\pm$40	    &842$\pm$33	    \\
\hline
\hline
\end{tabular}
\label{table:table2}
\end{table}

\subsection{Calculation of dipole moments}

The binary composition of these materials, and the inversion-asymmetric orientation of individual atoms at the unit cell creates an electric dipole in MLs that is oriented parallel to the longest lattice vector ($\mathbf{a}_1$), resulting in a piezoelectric response at 0 K.

But the orientation of these dipoles random at $T_c$ and the net magnitude of the electric dipole moment equal to zero. This hypothesis is demonstrated in Fig.~3(b) by computing the mean dipole over twelve average unit cells at a given temperature and without any additional optimization of the basis vectors. This was accomplished by feeding the instantaneous structural averages into the VASP code with PBE PAW pseudopotentials. In working with instantaneous averaged unit cells, we trade a spatial inhomogeneity by a temporal one. And as $T_c$ is approached, the relative orientation of the instantaneous average orientation among the atoms in the unit cell fluctuates, and with it the orientation of the dipole moment. The total moment decreases significantly as a result.

%



%

\end{document}